\documentclass[aps,twocolumn,prb,showpacs,preprintnumbers,superscriptaddress,amsmath,amssymb]{revtex4}
    \usepackage[pdftex]{graphicx} %%Grafiken und normales LaTeX
\usepackage[english]{babel}
\usepackage[ansinew]{inputenc}
\usepackage{blindtext}
\usepackage{natbib}
\usepackage{color}
\usepackage{dcolumn}% Align table columns on decimal point
\usepackage{bm}% bold math
\usepackage{ulem}
\newcommand{\co}[2]{\ifcase #1 \or #2 \fi}

\newif\ifnote

\begin{document}
%\ifpdf
%    \DeclareGraphicsExtensions{.pdf,.jpg,.png}
%\else
%    \DeclareGraphicsExtensions{.eps}
%\fi

\title{Modeling the linewidth dependence of coherent terahertz emission from intrinsic Josephson junction stacks in the hot-spot regime}
% Force line breaks with \\

\author{B. Gross}
\affiliation{
Physikalisches Institut and Center for Collective Quantum Phenomena in LISA$^+$,
Universit\"{a}t T\"{u}bingen,
D-72076 T\"{u}bingen,
Germany
}

\author{J. Yuan}
%\email{yuan.jie@nims.go.jp}
\affiliation{National Institute for Materials Science, Tsukuba 3050047, Japan}

\author{D.Y. An}
\affiliation{National Institute for Materials Science, Tsukuba 3050047, Japan}
\affiliation{Research Institute of Superconductor Electronics, Nanjing University, Nanjing 210093, China}

\author{M. Y. Li}
\affiliation{National Institute for Materials Science, Tsukuba 3050047, Japan}
\affiliation{Research Institute of Superconductor Electronics, Nanjing University, Nanjing 210093, China}

\author{N. Kinev}
\affiliation{Kotel'nikov Institute of Radio Engineering and Electronics, Russia}

\author{X. J. Zhou}
\affiliation{Research Institute of Superconductor Electronics, Nanjing University, Nanjing 210093, China}

\author{M. Ji}
\affiliation{National Institute for Materials Science, Tsukuba 3050047, Japan}
\affiliation{Research Institute of Superconductor Electronics, Nanjing University, Nanjing 210093, China}

\author{Y. Huang}
\affiliation{Research Institute of Superconductor Electronics, Nanjing University, Nanjing 210093, China}

\author{T. Hatano}
\affiliation{National Institute for Materials Science, Tsukuba 3050047, Japan}

\author{R.G. Mints}
\affiliation{The Raymond and Beverly Sackler School of Physics and Astronomy, Tel
Aviv University, Tel Aviv 69978, Israel}

\author{V. P. Koshelets}
\affiliation{Kotel'nikov Institute of Radio Engineering and Electronics, Russia}

\author{P.H. Wu}
\affiliation{Research Institute of Superconductor Electronics, Nanjing University, Nanjing 210093, China}

\author{H. B. Wang}
%\email{josephson.vortex@gmail.com}
\affiliation{National Institute for Materials Science, Tsukuba 3050047, Japan}
\affiliation{Research Institute of Superconductor Electronics, Nanjing University, Nanjing 210093, China}

\author{D. Koelle}
\affiliation{
Physikalisches Institut and Center for Collective Quantum Phenomena in LISA$^+$,
Universit\"{a}t T\"{u}bingen,
D-72076 T\"{u}bingen,
Germany
}

\author{R. Kleiner}
\affiliation{
Physikalisches Institut and Center for Collective Quantum Phenomena in LISA$^+$,
Universit\"{a}t T\"{u}bingen,
D-72076 T\"{u}bingen,
Germany
}

\date{\today}

\begin{abstract}
Recently it has been found that, when operated at large input power, the linewidth $\Delta f$ of terahertz radiation emitted from intrinsic Josephson junction stacks can be as narrow as some megahertz. In this high-bias regime a hot spot coexists with regions which are still superconducting. Surprisingly, $\Delta f$ was found to \textit{decrease} with increasing bath temperature. We present a simple model describing the dynamics of the stack in the presence of a hot spot by two parallel arrays of pointlike Josephson junctions and an additional shunt resistor in parallel. Heat diffusion is taken into account by thermally coupling all elements to a bath at temperature $T_b$. We present current-voltage characteristics of the coupled system and calculations of the linewidth of the radiation as a function of $T_b$. In the presence of a spatial gradient of the junction parameters critical current and resistance, $\Delta f$ deceases with increasing $T_b$, similar to the experimental observation. 
\end{abstract}

% PACS, the Physics and Astronomy Classification Scheme.


\maketitle

\section{Introduction}
\label{sec:intro}
Terahertz generation utilizing stacks of intrinsic Josephson junctions (IJJs) in the high-transition-temperature (high-T$_c$) cuprate Bi$_2$Sr$_2$CaCu$_2$O$_{8+\delta}$ (BSCCO) has become a major field of research, 
both in terms of experiment 
\cite{Ozyuzer07, Wang09a, Minami09, Kurter09, Gray09, Ozyuzer09, Guenon10, Wang10a, Tsujimoto10, Koseoglu11, Benseman11, Yamaki11, Yuan12, Li12, Tsujimoto12, Kakeya12, Tsujimoto12a, Turkoglu12, Oikawa13, Benseman13, Benseman13a}
%\cite{Kadowaki08,Ozyuzer09a,Kadowaki10}
and theory
\cite{Bulaevskii07, Koshelev08,Koshelev08b,Lin08,Krasnov09,Klemm09,Nonomura09,Tachiki09,Pedersen09,Hu09,Koyama09,Grib09,Zhou10,Krasnov10,Koshelev10, Savelev10,Lin10,Katterwe10,Yurgens11, Koyama11, TachikiT11,Slipchenko11,Krasnov11,Yurgens11b,Lin11b, Asai12,Asai12b,Zhang12,Lin12,Averkov12,Grib12,Gross12,Apostolov12,Liu13}; for reviews see Refs. \onlinecite{Hu10} and \onlinecite{Kashiwagi12}.
%\cite{Bulaevskii07, Koshelev08,Hu09,Koyama09,Grib09,Savelev10,Hu10,Lin10,Zhou10,Katterwe10, TachikiT11, Yurgens11b,Lin11b, Koyama11, Asai12,  Apostolov12}
% \cite{Wang10a,Guenon10,Yurgens11,Tachiki11}.
%
Typical IJJ stacks contain 500 -- 2000 IJJs and are either patterned as mesas on top of BSCCO single crystals, as Z-type all-superconducting structures \cite{Yuan12}, or as free-standing mesas sandwiched between gold electrodes \cite{Kashiwagi12,An13}. 
Emission frequencies are in the range 0.4 -- 1\,THz, with a maximum output power of several tens of $\mu$W emitted into free space \cite{Kashiwagi12,An13}. For arrays of several mesas even hundreds of $\mu$W have been achieved \cite{Benseman13a}.
Operated at a bath temperature $T_b$ well below $T_c$, there are two emission regimes. At moderate input power (``low-bias regime'') there is only little heating, and the temperature distribution in the mesa is roughly homogeneous and close to $T_b$. At high input power (``high-bias regime'') a hot spot\cite{Gurevich87} (an area heated to above $T_c$) forms inside the mesa, leaving the ``cold'' part of the mesa for terahertz generation via the Josephson effect.  
With respect to the linewidth $\Delta f$ of radiation one observes values of 0.5\,GHz or larger at low bias \cite{Kashiwagi12,Li12}. In the presence of a hot spot $\Delta f$ can be much lower, reaching values down to $\sim$ 20\,MHz \cite{Li12,An13}. The strong difference in $\Delta f$  at, respectively, high and low bias strongly indicates that -- in addition to cavity resonances which seem to play an important role for synchronization both at high and low bias \cite{Ozyuzer07,Benseman11,Wang10a} -- the hot spot also is essential for synchronization. 
Further, it was found that $\Delta f$ \textit{decreases} with increasing $T_b$ \cite{Li12}. This behavior is quite unusual for any Josephson junction based oscillator. 

Thus, there is a clear need to investigate the dynamics of Josephson junctions in the presence of strong heating. Temperature distributions in IJJ mesas  have been simulated in Refs. \onlinecite{Yurgens11,Yurgens11b,Gross12} by solving the 3D heat diffusion equations in the \textit{absence} of Josephson currents. It has been shown that the peculiar temperature dependence of the BSCCO $c$-axis resistance is the main ingredient being responsible for hot-spot formation \cite{Gross12}.  Following Ref. \onlinecite{Spenke36b}, in Ref. \onlinecite{Gross12} a simple two-resistor model with thermal coupling to a bath was presented, which is based on the temperature dependent BSCCO $c$-axis resistance and the $c$-axis thermal conductance to describe hot-spot formation and the shape of the current-voltage characteristic of the IJJ stacks. In the present paper we adopt this approach to include the effect of Josephson currents.  

The model, presented in Sec. \ref{sec:model}, starts with a stack of $N$ Josephson junctions. As a first step, we assume that all junctions oscillate in phase, acting as a single giant Josephson junction. Subsequently, we split the giant junction into  $M = N/p$ segments in $c$-direction. Here, $p$ is a prime factor of $N$. In each segment the junctions are assumed to behave identical and are described by the resistively and capacitively shunted junction (RCSJ) model \cite{Stewart68,McCumber68}. 
For both the giant junction and the segmented junctions the stack is split in lateral direction into two parts at, respectively, temperatures $T_1$ and  $T_2$, to be calculated from a balance between the heat generation in the two parts and the vertical heat transfer to a bath. Simulations by Yurgens \cite{Yurgens11} showed that a distributed network of resistors and capacitors representing the interior of the hot spot can synchronize an array of (pointlike) Josephson junctions. Thus, to provide potential phase synchronization, as an additional element to the segmented junctions a resistor at temperature $T_2$ is attached across the whole array, representing the interior of the hot spot. 

As we will show in Sec. \ref{sec:results} our model indeed allows for a linewidth of the radiation which decreases with increasing bath temperature. A necessary requirement is that the junction parameters have a gradient in critical current and resistance, representing the finite slope of the mesa edges \cite{Benseman11}.

\section{Model}
%In this section we first introduce the simple model of a stack of identical junctions and then continue with the more complex model of a stack split into M segments and shunted by a resistor. In the third part we discuss the parameters used in our simulations.   

\label{sec:model}
\subsection{Stack of identical junctions}
\label{sec:model1}
We consider a stack of $N$ intrinsic Josephson junctions, each junction described within the RCSJ model, in combination with the time-dependent heat-diffusion equation taking into account self heating in the IJJ stack. 
%Under these assumptions the Josephson dynamics in the whole stack can be described by an equation which has the same form as for a single junction within the RCSJ model. 
The parameters resistance and critical current, as well as the Nyquist noise arising from the resistors, are temperature dependent. The electrical power generated by the Josephson junctions in the resistive state serves as input to the heat-diffusion equations to calculate the temperature of the stack. 
We split the stack laterally into two parts which for convenience we assume to have equal size. 
%, as indicated in Fig. \ref{fig:homstack}.  
Thus each junction in the stack consists of two parts connected in parallel. Each sub-junction is described by a parallel connection of a Josephson element, a resistor, a capacitor, and a noise source. We further neglect the resistance of the in-plane parallel wires (electrodes) connecting the two parts 
\footnote{Near the hotspot a substantial amount of current can, in principle, flow as a resistive in-plane current causing substantial in-plane electric fields. However, there is typically a low-resistance gold layer on top of the mesa, which homogenizes the current injected into the mesa, preventing strong in-plane currents in the stack itself. The precise impact of in-plane currents on, e. g. the shape of the hot spot and the Josephson dynamics can only be answered from explicit simulations using 1D or 2D coupled sine-Gordon equations.}
.
Then, all circuit elements carry the same voltage $U_n$ which, using the second Josephson relation, transforms into $\dot{\gamma}_{1,n}=\dot{\gamma}_{2,n} =2\pi U_n/\Phi_0$. $\Phi_0$ is the flux quantum. The first index on the Josephson phase differences $\gamma_{k,n}$  labels the two parts of the junction and $n$ = 1..$N$ labels the junctions. 
We further assume that no magnetic flux threads the loop formed between the two parts. Then, the Josephson phase differences $\gamma_{1,n}$ and $\gamma_{2,n}$  are equal, $\gamma_{1,n} = \gamma_{2,n} \equiv \gamma_{n}$.  
Under these assumptions the electrical part of the circuit reads
\begin{equation}
\label{eq:hom_RSJ_i}
\begin{split}
I = \,\, & \frac{2\pi (C_{1,n}+C_{2,n})}{\Phi_0}\ddot{\gamma}_{n} \\
		& + \frac{2\pi}{\Phi_0}[\frac{1}{R_{1,n}(T_1)}+\frac{1}{R_{2,n}(T_2)}]\dot{\gamma}_{n} \\
    & + [I_{c1,n}(T_1)+I_{c2,n}(T_2)]\sin\gamma_n \\
		& + I_{N1,n}(T_1) + I_{N2,n}(T_2).
\end{split}
\end{equation}
where $C_{1,n}$, $C_{2,n}$, $R_{1,n}$, $R_{2,n}$, $I_{c1,n}$ and $I_{c2,n}$ are the junction capacitances, resistances and critical currents, with index $n$ = 1..$N$.  
We assume that the junction resistances and critical currents are temperature dependent. For convenience, we assume that the capacitances do not depend on temperature.
We further assume that all sub-junctions in part 1 of the stack are at temperature $T_1$ while the junctions in part 2 are at temperature $T_2$. This is justified from calculations of the heat-diffusion equations in the absence of Josephson currents\cite{Yurgens11}. 
%
%Figure Homogeneous stack %%%%%%%%%%%%%%%%%%%%%%%%%%%%%%%%%%%%%%%%%%%%%%%%%%%%%%%%%%
\begin{figure}[tb]
\includegraphics[width=\columnwidth,clip]{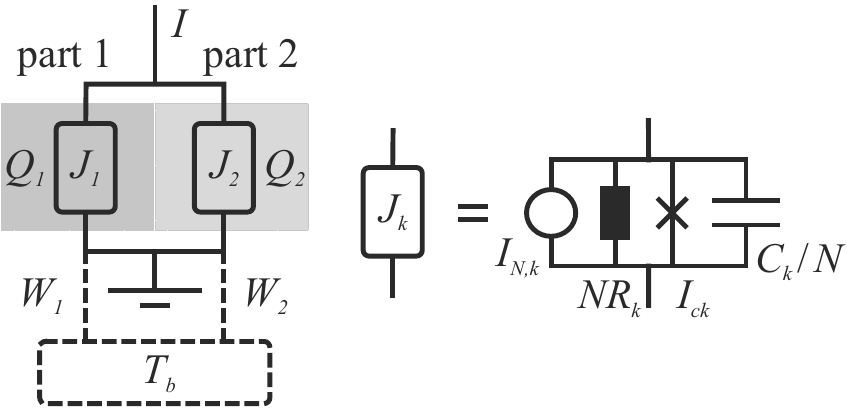}
\caption{A ``giant" intrinsic Josephson junction, laterally split into two parts that can be at different temperatures $T_1$ and $T_2$. Solid lines indicate electrical circuit, dashed lines thermal circuit. Heat transfer $W_k$ of the two parts is only to the bath but not between the two parts, $k$ = (1,2).
See Eq. (\ref{eq:hom_RSJ_i}) and the corresponding text for a discussion how individual junctions are electrically connected to form the giant junction. The scheme can also be seen in Fig. \ref{fig:Segmentstack} when interpreting the junctions $J_{k,m}$ in this graph as individual junctions laterally split into two parts.
}
\label{fig:homstack}
\end{figure}
%Figure Setup end %%%%%%%%%%%%%%%%%%%%%%%%%%%%%%%%%%%%%%%%%%%%%%%%%%%%%%
%

We now assume that all Josephson junctions oscillate in-phase, and sum up Eq. (\ref{eq:hom_RSJ_i}) over all $N$ junctions. By assumption all $\gamma_{n} = \gamma$ are equal, and we further assume that also the junction critical currents, capacitances and resistances do not depend on $n$. This yields  
\begin{equation}
\label{eq:hom_RSJ}
\begin{split}
I = \,\, & \frac{2\pi (C_{1}+C_{2})}{N\Phi_0}(N\ddot{\gamma}) \\
		& + \frac{2\pi}{\Phi_0}[\frac{1}{NR_{1}(T_1)}+\frac{1}{NR_{2}(T_2)}](N\dot{\gamma}) \\
    & + [I_{c1}(T_1)+I_{c2}(T_2)]\sin\gamma \\
		& + \frac{1}{N} \sum_{n=1}^N \left[ I_{N1,n}(T_1) + I_{N2,n}(T_2)\right],
\end{split}
\end{equation}
where $2\pi N\dot{\gamma}/\Phi_0$ is the voltage $NU$ across the whole stack. $C_{k}/N$  and $NR_{k}$ are the total capacitance and resistance of the two segments. For the noise currents 
\begin{equation}
\label{eq:noise_currents}
I_{Nk} = \frac{1}{N} \sum_{n=1}^NI_{Nk,n}
\end{equation}
%$I_{Nk,n} = \frac{1}{N} \sum_{n=1}^NI_{Nk,n}$ 
we assume a white spectral power density 
\begin{equation}
S_{I,k}=4k_B \frac{T_k}{NR_k}
\end{equation}
%
%$S_{I,k}$=4k$_B$ $T_k/NR_k$, 
with $k$ = (1,2) and the Boltzmann constant $k_B$. Thus, the stack behaves as a giant junction, as sketched in Fig. \ref{fig:homstack}.

Note that one or even both parts of the stack can be above the critical temperature $T_c$. Then, $I_{c1}$, $I_{c2}$ or both are zero, while the other parameters remain finite. For $I_{c1} + I_{c2} = 0$ Eq. (\ref{eq:hom_RSJ}) still is solvable, and $\Phi_0\dot{\gamma}/2\pi$ acts just as a somewhat unusual expression for the voltage $U$. The more critical term  $(I_{c1} + I_{c2})\sin\gamma$ involving the phase $\gamma$ -- the concept which is not defined above $T_c$ -- has disappeared. 

Assuming that there is no (in-plane) heat transfer between the two parts of the stack the thermal part of our system is given by
\begin{equation}
\label{eq:hom_Therm1}
\tilde{C}_1\dot{T}_1=\frac{U^2}{R_1(T_1)}-\frac{K_1}{N}(T_1-T_b)
\end{equation}
and
\begin{equation}
\label{eq:hom_Therm2}
\tilde{C}_1\dot{T}_2=\frac{U^2}{R_2(T_2)}-\frac{K_2}{N}(T_2-T_b).
\end{equation}
$\tilde{C}_1$ and $\tilde{C}_2$ are the heat capacitances per junction of the two parts, which below we take equal and temperature independent. $Q_1 =U^2/R_1$ and $Q_2 = U^2/R_2$ represent the Joule heating power per junction generated by the two parts of the stack. $K_1$ and $K_2$ are the $c$-axis thermal conductances of the stack to the bath at temperature $T_b$.
Below we will assume temperature independent constants $K_1 = K_2 = K$. 

Note that we did not include an in-plane heat transfer $K_{12}$. The BSCCO in-pane conductivity $\kappa_{ab}$ is roughly five times bigger than the out-of-plane conductivity $\kappa_{c}$. On the other hand, a typical mesa is $\sim$1\,$\mu$m thick and $\sim$300\,$\mu$m long. $K_1$ and $K_2$ are inversely proportional to the mesa length while $K_{12}$ is inversely proportional to its thickness. Thus, the geometric aspect ratio overwhelms the anisotropy in the heat conductances. On the next level one should further consider the temperature distribution in the base crystal, which is some 10-50\,$\mu$m thick and some 500\,$\mu$m long and has bottom cooling. Also, one should allow that the size of the hot spot is variable. 
Continuing along this line one realizes that the next iteration is the ``1D'' model, as used in Ref. \onlinecite{Gross12}, which is out of the scope of the present model.

We next write Eqs. (\ref{eq:hom_RSJ}) -- (\ref{eq:hom_Therm2}) in a normalized form, using the 4.2\,K values of the various parameters as reference. Currents are measured in units of $I_{c0} =  I_{c1}($4.2\,K$)+I_{c2}($4.2\,K$)$, resistances in units of the total resistance of one IJJ, $R_0 = [R_{1}($4.2\,K$)^{-1}+R_{2}($4.2\,K$)^{-1}]^{-1}$, capacitances in units of the total capacitance $C= C_1+C_2$ per junction, voltages in units of $I_{c0}R_0$ and time in units $\tau = \Phi_0/2\pi I_{c0}R_0$. The spectral density of the normalized noise current $i_{Nk}$ is
\begin{equation}
s_{i,k} = 4\frac{\Gamma_0}{N}\frac{T}{4.2\,K}\frac{R_0}{R_k(T)},
\end{equation}
%
%4($\Gamma_0/N$)$\cdot$($T$/4.2\,K)/($R_k(T)/R_0$), 
with  $\Gamma_0 = 2\pi k_B\cdot$4.2\,K/($I_{c0}$$\Phi_0$).  Using  
\begin{equation}
\beta_c=\frac{2\pi C I_{c0}R_{0}^2}{\Phi_0} 
\end{equation}
%
%$\beta_c=2\pi C I_{c0}$$R_{0}^2$/$\Phi_0$ 
we obtain the normalized form of Eq. (\ref{eq:hom_RSJ})  as

\begin{equation}
\label{eq:hom_RSJn}
\begin{split}
i=& \beta_c[c_1+c_2]\ddot{\gamma}+[\frac{1}{r_1(T_1)}+\frac{1}{r_2(T_2)}]\dot{\gamma}\\
  &+[i_{c1}(T_1)+i_{c2}(T_2)]\sin\gamma+i_{N1}(T_1)+i_{N2}(T_2),\\&
\end{split}
\end{equation}
with $i = I/I_{c0}$, $i_{ck}(T_k)=I_{ck}(T_k)/I_{c0}$, $r_k(T_k)=R_k(T_k)/R_0$,  $c_k=C_k/C$, $i_{N,k}=I_{N,k}/I_{c0}$. Eq.  (\ref{eq:hom_RSJn}) is close to the standard form of the (single junction) RCSJ equation. Note, however, that the parameters $r_1,r_2,i_{c1}$ and $i_{c2}$ are time dependent through the time dependence of the temperatures $T_1$ and $T_2$. 

In normalized form Eqs. (\ref{eq:hom_Therm1}) and (\ref{eq:hom_Therm2}) read
\begin{equation}
\label{eq:hom_Therm1n}
\tilde{c}_1\dot{T}_1=\frac{\dot{\gamma}^2}{r_1(T_1)}-k_1(T_1-T_b)
\end{equation}
and
\begin{equation}
\label{eq:hom_Therm2n}
\tilde{c}_2\dot{T}_2=\frac{\dot{\gamma}^2}{r_2(T_2)}-k_2(T_2-T_b),
\end{equation}
with $\tilde{c}_k=2\pi\tilde{C_k}/\Phi_0I_{c0}$ and $k_k=K_k/NI_{c0}^2$$R_0$. The quantities $\tilde{c}_k$ and $k_k$ are in units of 1/K and the temperatures $T_1$ and $T_2$ are still dimensioned. We have also used the normalized second Josephson relation $u=\dot{\gamma}$, where $u$ is the normalized voltage per junction.

\subsection{Segmented stack}
\label{sec:segments}
%
%Figure Homogeneous stack %%%%%%%%%%%%%%%%%%%%%%%%%%%%%%%%%%%%%%%%%%%%%%%%%%%%%%%%%%
\begin{figure}[tb]
\includegraphics[width=\columnwidth,clip]{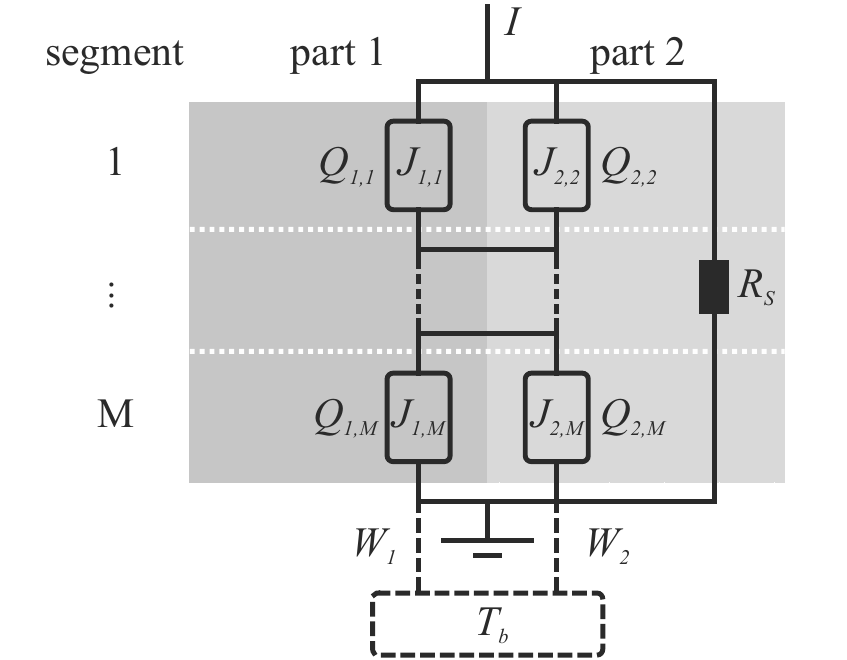}
\caption{A stack of intrinsic Josephson junctions consisting of $M$ segments as a generalization of the circuit shown in Fig. \ref{fig:homstack}. Each segment is laterally split into two parts that can be at different temperatures $T_1$ and $T_2$. A shunt resistor $R_s$ at temperature $T_2$, representing the inner part of the hot spot, is mounted across the whole stack. 
}
\label{fig:Segmentstack}
\end{figure}
%Figure Setup end %%%%%%%%%%%%%%%%%%%%%%%%%%%%%%%%%%%%%%%%%%%%%%%%%%%%%%
%
We next turn to the actual model used for our calculations. A schematic is shown in Fig. \ref{fig:Segmentstack}. Using the same normalizations it is straightforward to arrive at the equations
\begin{equation}
\label{eq:segment_RSJn}
\begin{split}
i=\, & \beta_c[c_1+c_2]\ddot{\gamma}_{m}+[\frac{1}{r_1(T_1)}+\frac{1}{r_2(T_2)}]\dot{\gamma}_{m}\\
     & +[i_{c1,m}(T_1)+i_{c2,m}(T_2)]\sin\gamma_{m}+i_{N1,m}(T_1)\\
     & +i_{N2,m}(T_2)+i_{Ns}(T_2)+\frac{N}{Mr_s(T_2)}\sum_{m=1}^M\dot{\gamma}_{m},\\
\end{split}
\end{equation}
\begin{equation}
\label{eq:segment_Therm1n}
\tilde{c}_1\dot{T}_1=\frac{1}{M}\sum_{m=1}^M\frac{\dot{\gamma}_{m}^2}{r_{1,m}(T_1)}-k_1(T_1-T_b)
\end{equation}
and
\begin{equation}
\label{eq:segment_Therm2n}
\begin{split}
\tilde{c}_2\dot{T}_2 = \,& \frac{1}{M}\sum_{m=1}^M\frac{\dot{\gamma}_{,m}^2}{r_{2,m}(T_2)}+\frac{N}{r_s(T_2)}(\frac{1}{M}\sum_{m=1}^M\dot{\gamma}_{m})^2\\
											   & -k_2(T_2-T_b).
\end{split}
\end{equation}
In Eq. (\ref{eq:segment_RSJn}) the index $m$ runs from 1..$M$. 
The last term in Eq. (\ref{eq:segment_RSJn}) represents the normalized current through the resistor $r_s$. This resistor, which we assume to have the same temperature $T_2$ as array 2, generates a noise current $i_{Ns}$ with spectral power density 
\begin{equation}
s_s=4\Gamma_0\frac{T_2}{4.2\,K}\frac{1}{r_s(T_2)}. 
\end{equation}
%
%$4\Gamma_0(T_2/4.2$\,K)/$r_s(T_2)$. 
The spectral power densities of the noise currents $i_{Nk,m}(T_k)$ are given by 
\begin{equation}
s_{i,k,m}= 4\Gamma_0\frac{T_k}{4.2\,K}\frac{M}{N r_{k,m}(T_k)}.
\end{equation}
%$4M\Gamma_0(T_k/4.2$\,K)/$N r_{k,m}(T_k)$.
%
The second term in Eq. (\ref{eq:segment_Therm2n}) represents ohmic heating in the resistor $r_s$.
For $R_s$ we will assume the same temperature dependence as for the other resistors. Unless stated differently, throughout the paper we will also assume that $r_s/N = r_2= M^{-1}\Sigma_{m=1}^Mr_{2,m}$, i. e. half of the in-plane area of the ``hot'' part of the stack is formed by the shunt resistor. The hot area itself shall cover half of the junction area, i. e. $r_1^{-1}=r_2^{-1}+(r_s/N)^{-1}$ , with $r_1= M^{-1}\Sigma_{m=1}^Mr_{1,m}$ . 
We will, unless stated differently, also use identical 4.2\,K values of the critical currents and resistances of all segments. 

In the limit $R_s \rightarrow\infty$ the last term in Eq. (\ref{eq:segment_RSJn}) disappears and the $M$ segments are uncoupled except for a parametric coupling introduced through the time dependence of temperatures $T_1$ and $T_2$, as calculated in Eqs. (\ref{eq:segment_Therm1n}) and (\ref{eq:segment_Therm2n}). In principle, this coupling can introduce phase lock between the segments (we have tested this), however, only if the thermal part of the circuit becomes unrealistically fast, i. e. the $\tilde{c}_k$ become very small. 
It has been shown in Ref. \onlinecite{Yurgens11} that a distributed network of resistors and capacitors modeling the hot spot can provide phase lock. In such a network there are not only current paths which connect adjacent junctions but also paths which connect more distant junctions.  In our lumped circuit model the most simple synchronizing element representing this is the resistor $R_s$ in parallel to the two junction arrays.
Note that we have omitted this resistor in Fig. \ref{fig:homstack}. Here, this resistor would just add in parallel to the resistor $R_2$, yielding no new information.
Further, we could have also chosen a model where only one junction array (at temperature $T_1$) is present and is shunted by a resistor at temperature $T_2$, representing the hot spot. In fact we have studied this model. It however turns out that over a wide range of currents $i >>  1$ the array can be multistable, allowing both for a resistive and a zero voltage state for each junction. The reason is that the actual current through the array is well below the critical current of the various junctions even for $i >>  1$. This situation is not observed in experiment, at least as long as a single stack with a hot spot is considered. The model shown in Fig. \ref{fig:Segmentstack} thus seems to be the minimal model to describe both heating effects in an IJJ stack and phase synchronization phenomena.

\subsection{Choice of parameters}
\label{sec:params}
For an intrinsic Josephson junction at $T$ = 4.2\,K one typically finds $I_{c0}R_0 \sim 15$\,mV, corresponding to a characteristic frequency $f_{c0} = I_{c0}R_0/\Phi_0 \sim 7.5$\,THz. The Josephson plasma frequency $f_{pl0} = f_{c0}/\sqrt{\beta_c} < 150$\,GHz and thus $\beta_c >$ 2500. Numerically, so large numbers cause instabilities and may even not be realistic due to additional (high frequency) damping mechanisms. Thus, in the calculations discussed below we use $\beta_c$ = 200. 
For the large mesas used for terahertz generation one typically has $I_{c0} \sim 30$\,mA, leading to a ``characteristic power'' $I_{c0}^2 R_0 \sim 0.5$\,mW per junction. 
In Ref. \onlinecite{Gross12} $K \sim 6 \cdot 10^{-4}$\,W/K has been used to reasonably fit the heating properties of  a $300 \,\times\, 50\, \mu$m$^2$  mesa with $N \sim$ 670 IJJs. This leads to $k \sim 1.8 \cdot 10 ^{-3}$/\,K. Below we will use $k$ = 10$^{-3}$/\,K. 
The effective heat capacitance of the whole mesa is hard to estimate because of the various contacting and glue layers. It is also strongly temperature dependent \cite{Suzuki10}. However, later on we will be interested in situations where the stack has reached a constant temperature so that the exact value does not matter very much. For simplicity, using a specific heat capacitance of 50\, J/m$^3$K, which is a typical number for $T \sim 50$\,K, we obtain $\tilde{c} \sim 100$/\,K. We use this value for the calculations shown below.
For $I_{c0} = 30$\,mA we further obtain  $\Gamma_0 = 5.9\cdot10^{-6}$, yielding $\Gamma_0/N \sim 10^{-8}$. In Sec. \ref{sec:results} we will see that over a wide temperature range the normalized linewidth $\Delta f/f_{c0}$ of the Josephson oscillations is on the order of $\Gamma_0/N$, if the junctions are phase locked.  The integration time used for the calculation should be well above the reciprocal linewidth to resolve the line. This is too time consuming for a realistic value of  $\Gamma_0/N$. Below we thus use $\Gamma_0M/N = 10^{-4}$  to make calculations feasible.

For the temperature dependence of $I_c$, for $T < T_c$ = 80\,K we use the parabolic form 
\begin{equation}
\label{eq:IcT}
I_c(T)= I_c(0) [1-(T/T_c)^2].
\end{equation}
For $T > T_c$ $I_c(T)$ = 0.
%
%Figure R(T) %%%%%%%%%%%%%%%%%%%%%%%%%%%%%%%%%%%%%%%%%%%%%%%%%%%%%%%%%%
\begin{figure}[tb]
\includegraphics[width=1\columnwidth,clip]{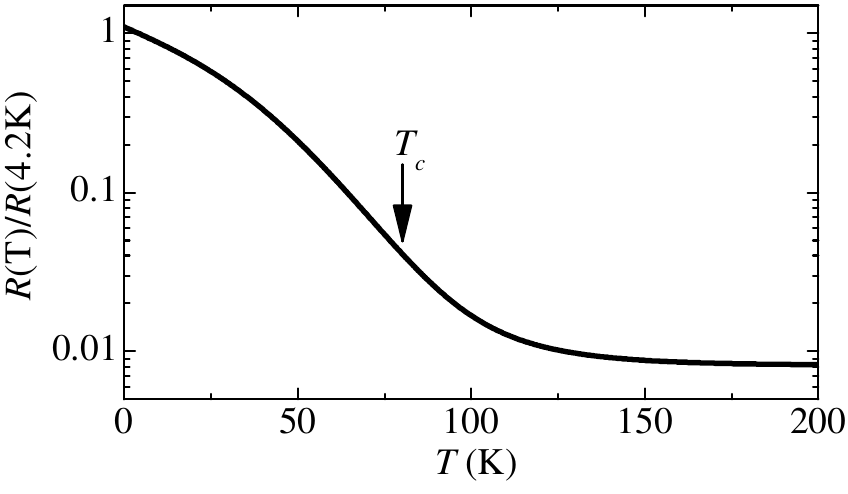}
\caption{Temperature dependence of $c$-axis resistance, as used for calculations, Eq. (\ref{eq:RT}). $T_c$ is the critical temperature.
}
\label{fig:RT}
\end{figure}
%Figure R(T) end %%%%%%%%%%%%%%%%%%%%%%%%%%%%%%%%%%%%%%%%%%%%%%%%%%%%%%
%
For the $c$-axis resistance we use a somewhat complex expression, to be normalized to $R_0$:
\begin{equation}
\label{eq:RT}
\begin{split}
R(T)\,\, & = 6 [\exp(-T/22\,\rm{K}) \\
				 & + \exp(-T^2/1900\,\rm{K}^2)] +0.09.
\end{split}
\end{equation}
This expression, shown in Fig. \ref{fig:RT} is an approximate fit to the BSCCO $c$-axis $R(T)$ curve used in Ref. \onlinecite{Gross12}.
For temperatures above the transition temperature $T_c$ the experimental  $R(T)$ can be measured directly, below $T_c$ it can be 
estimated either extrapolating the resistive branches of the current-voltage characteristic (IVC) to zero current \cite{Yurgens11, Benseman13} or by adjusting it so that measured over-heated IVCs are reproduced. 

\section{Results}
\label{sec:results}
%Figure IVCs %%%%%%%%%%%%%%%%%%%%%%%%%%%%%%%%%%%%%%%%%%%%%%%%%%%%%%%%%%
\begin{figure}[tb]
\includegraphics[width=1\columnwidth,clip]{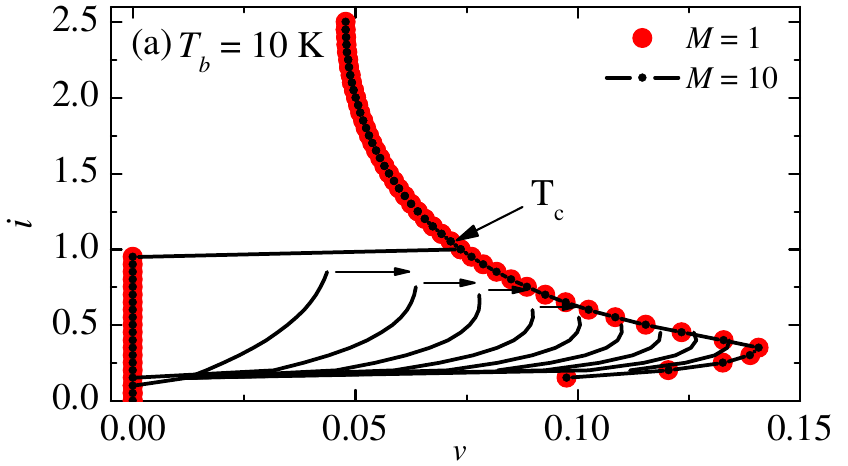}
\includegraphics[width=1\columnwidth,clip]{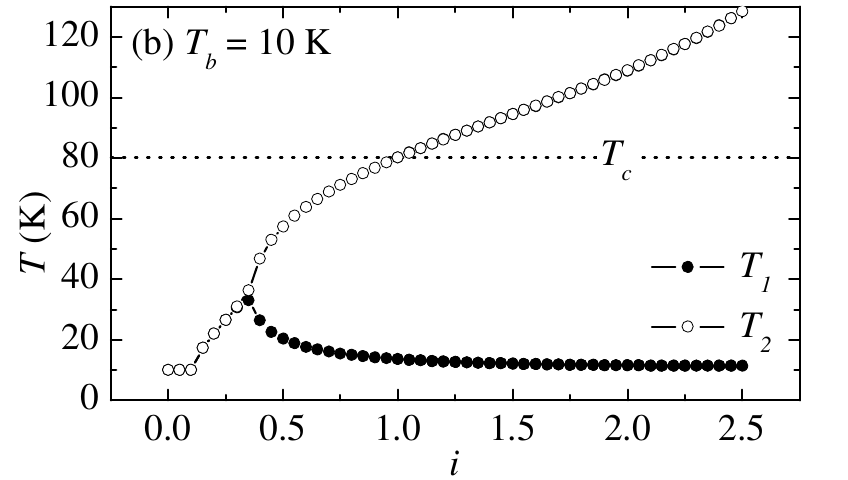}
\caption{(Color online) (a) Current-voltage characteristic for $T_b$ = 10\,K, $M$ = 1 and $M$ = 10. (b) Temperatures $T_1$ and $T_2$ of the two parts of the IJJ stack vs current $i$ for $T_b$ = 10 K. $T_c$ denotes the temperature where the hot part has reached the superconducting transition.
}
\label{fig:IVC}
\end{figure}
%Figure IVC end %%%%%%%%%%%%%%%%%%%%%%%%%%%%%%%%%%%%%%%%%%%%%%%%%%%%%%
%
Let us first look at IVCs, as calculated from Eqs. (\ref{eq:segment_RSJn}) -- (\ref{eq:segment_Therm2n}) 
using a 5th order Runge-Kutta method. An IVC is typically calculated by starting with $i = 0$ and initial conditions $\gamma_{1,m} = \dot{\gamma}_{1,m}=0$, $T_1 = T_b$ and $T_2 = 1.01T_b$ and later on increasing $i$ by some step $\Delta i$, keeping the values of  $\gamma_{1,m}$, $\dot{\gamma}_{1,m}$, $T_1$ and $T_2$ from the previous step as initial conditions. Having reached some maximum value of $i$ the current is decreased back to 0.
To calculate $u$ for a given $i$ we choose a time step  $\Delta t=0.2 /r$, where $r = [r_1^{-1}+r_2^{-1}+(r_s/N)^{-1}]^{-1}$ is the normalized resistance per junction, and then let the system evolve for 100.000 time steps to reach a stationary state.  This step is repeated until the temperatures $T_1$ and $T_2$ are stable within 1$\%$. We then take data over 100.000 time steps to calculate the average voltage $v$ per junction. 

\begin{figure}[tb]
\includegraphics[width=1\columnwidth,clip]{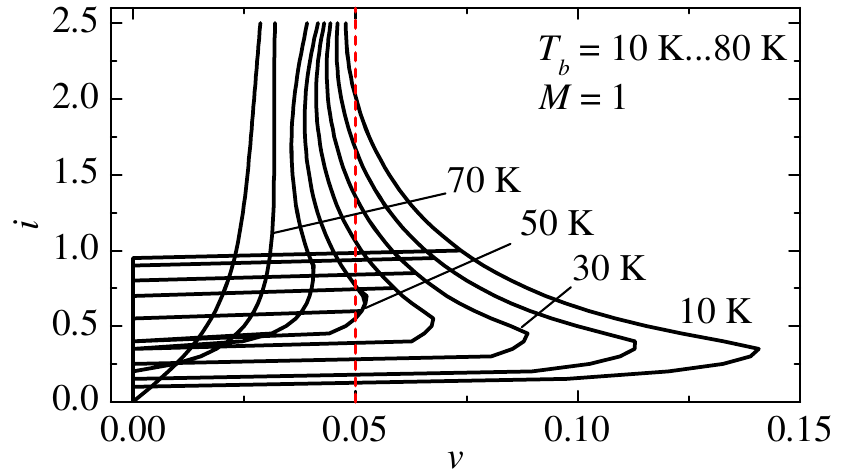}
\caption{(Color online) Current-voltage characteristics for $M$ = 1 and bath temperatures between 10 K and 80 K, in steps of 10 K.}
\label{fig:IVCs}
\end{figure}

In Fig. \ref{fig:IVC}(a) we compare IVCs, as calculated at $T$ = 10\,K for $M = 1$ and for $M = 10$. Both curves nearly coincide, which is due to the normalizations used (in fact, also IVCs for $r_s \rightarrow\infty$ would lie on top of the IVCs shown). For the $M$ = 10 case we also traced out the 9 inner branches $n=1..9$, where $n$ of the segments are in the resistive state while $M-n$ segments are in the zero voltage state. The branches are traced out by  choosing initial conditions $\dot{\gamma}_{1,m}=ri$ for the segments desired to be resistive, while using $\dot{\gamma}_{1,m}=0$ for the other segments. The IVCs shown in Fig. \ref{fig:IVC}(a) closely resemble experimental data \cite{Tsujimoto12a}. The maximum voltage $v \approx 0.14$, corresponding to $V \approx 2$ mV in dimensioned units, is reached for $i \approx 0.35$. Here, $T_1 \approx T_2 \approx$ 32\,K, compare Fig. \ref{fig:IVC} (b). For larger currents $T_2$ becomes larger than $T_1$, e.g. reaching $T_c = 80$\,K at $i\approx 1$. Here, $T_1 \approx 15$\,K. Fig. \ref{fig:IVCs} shows IVCs for $M = 1$, for bath temperatures between 10\,K and 80\,K. Also these IVCs closely resemble experimental curves.

%Figure T1T2 %%%%%%%%%%%%%%%%%%%%%%%%%%%%%%%%%%%%%%%%%%%%%%%%%%%%%%%%%%
%Figure T1T2 end %%%%%%%%%%%%%%%%%%%%%%%%%%%%%%%%%%%%%%%%%%%%%%%%%%%%%%
%
The main purpose of our calculations is to investigate the linewidth $\Delta f$ of emission as a function of bath temperature. In the experiments of Ref. \onlinecite{Li12}, $\Delta f$ vs. $T_b$ has been determined at a fixed frequency of $f\approx$ 0.62\,THz, corresponding to $V \approx$ 1.3\,mV or $v \approx$ 0.08.  For our simulation we have chosen a somewhat smaller value, $v$ = 0.05. The (red) dashed line shown in Fig. \ref{fig:IVC}(b) indicates bias points at various bath temperatures where the same voltage $v$ = 0.05 is realized. Note that for $T_b$ = 50\,K the (red) dashed line intersects the IVC both at high bias, i. e. in the region of negative differential resistance and at low bias, at $i\approx 0.55$. 
For this value of $v$ we can determine ``emission'' spectra at temperatures between 10\,K and 50\,K. ``Emission'' spectra are calculated by recording the voltage $u$ across the stack over a reasonably long time of typically $5\cdot10^5$ time units, taking a Fourier transform and averaging the resulting power spectrum up to 200 times. 

For a pointlike Josephson junction with resistance $R$ at \textit{fixed} temperature $T$ and no back-action of temperature fluctuations to the junction parameters the linewidth of radiation is given by \cite{Larkin68,Dahm69}
\begin{equation}
\label{Eq:delta_f}
\Delta f = \frac{4\pi k_B Tr^2}{\Phi_0^2R},
\end{equation}
where $r$ is the differential resistance at the bias point. 
Using $r = R$, which is a  good approximation as long as the bias current is well above the critical current, one obtains $\Delta f =4\pi k_BT R/\Phi_0^2$, and, with the normalization of frequencies to $f_{c0}$, a dimensionless linewidth 
\begin{equation}
\Delta f = 2\Gamma_0\frac{R}{R_0}\frac{T}{4.2\,K}.
\end{equation}
%
%$\Delta f = 2\Gamma_0[R/R_0][T/4.2\,K]$. 
Note that for large BSCCO stacks we cannot determine the differential resistance $r$ from measured IVCs, because the temperatures of both the cold and the hot part vary strongly with the bias current. Still, one may use $R = V/I$ to obtain the resistance at a given bias point. 
For the case of hot and cold regions in parallel one can, following Ref. \onlinecite{Larkin68}, define an effective temperature via 
\begin{equation}
\label{eq:T_eff}
T_{\rm{eff}} = R_{\rm{eff}}\left[\frac{T_1}{R_1}+\frac{T_2}{R_2}+\frac{T_2N}{R_s}\right],
\end{equation}
where $R_{\rm{eff}} = R_0\cdot v/i $ is the resistance of the 3 parts of the stack connected in parallel. The dimensionless linewidth in this case is 
\begin{equation}
\label{eq:f_0}
\Delta f_0=2\Gamma_0 \frac{R_{\rm{eff}}}{R_0} \frac{T_{\rm{eff}}}{4.2\,K},
\end{equation}
in units of the 4.2 K characteristic frequency. Since the cold part at temperature $T_1$ has a high resistance this roughly reduces to $T_{\rm{eff}} = T_2$ and $\Delta f_0 = 2\Gamma_0 (v/i) [T_2/4.2\,K]$.  In our case $T_2$ is of order 100--130\,K in the high bias regime and, thus, the main change in $\Delta f_0$ comes from the factor $v/i$ which, according to 
Fig. \ref{fig:IVC}(b), increases from about 0.025 at $T_b = 10$\,K to 0.066 at $T_b = 50$\,K. Not very surprisingly we obtain a linewidth which increases with increasing $T_b$.  

Also note that  $(v/i) [T_2/4.2\,K]$ is roughly of order unity and thus  $\Delta f_0$ is of order $2\Gamma_0$ when we neglect back-actions of the temperature fluctuations to the junction stack. 
Performing simulations with fixed, i. e. time independent values of $T_1$ and $T_2$ we have tested the above relation for $\Delta f_0$ and found very good agreement. 
By contrast, including back-actions, we find in simulations that $\Delta f$ can differ from $\Delta f_0$. Particularly, at high current (relative to the critical current at given temperature) and for small values of $M$ it can become significantly lower than $\Delta f_0$. 
%
%Figure Emi1 %%%%%%%%%%%%%%%%%%%%%%%%%%%%%%%%%%%%%%%%%%%%%%%%%%%%%%%%%%
\begin{figure}[tb]
\includegraphics[width=1\columnwidth,clip]{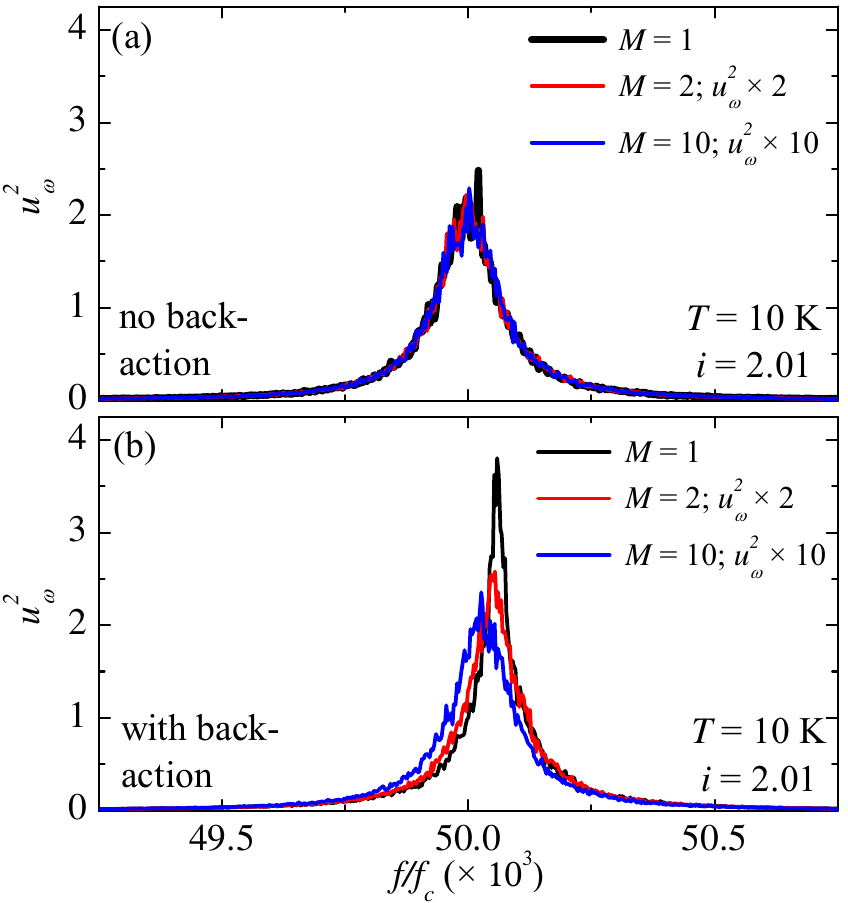}
\caption{(Color online) Fourier transforms (power) $u^2_\omega$ of the voltage $u$ across the stack vs. $f/f_c$ for $T_b$ = 10\,K, $i = 2.01$ and $r_s \rightarrow\infty$ for (a) time-independent temperatures $T_1$ and $T_2$ and (b) for the thermally coupled circuit.
}
\label{fig:emi1}
\end{figure}
%Figure Emi1 end %%%%%%%%%%%%%%%%%%%%%%%%%%%%%%%%%%%%%%%%%%%%%%%%%%%%%%
%
Fig. \ref{fig:emi1} shows the effect for $T_b$ = 10\,K, $i = 2.01$ and $M$ = 1, 2 and 10. In the graphs the power $u_\omega^2$ is plotted vs. $f/f_c$. For these simulations we have used $r_s \rightarrow\infty$. The Fourier spectra of Fig. \ref{fig:emi1} (a) have been calculated for time-independent values of $T_1$ = 11.5\,K and $T_2$ = 109\,K while for the curves in Fig. \ref{fig:emi1} (b) the coupled Eqs. (\ref{eq:segment_RSJn}) -- (\ref{eq:segment_Therm2n}) have been used. In both figures the Fourier spectra  for given $M$ are multiplied with $M$. The curves of Fig. \ref{fig:emi1} (a) are for uncoupled segments. Then, for the normalization used, one expects $\Delta f$ to be independent of $M$. The amplitude $u_\omega^2$ should decrease $\propto M^{-1}$, since the voltages $u$ for the $M$ segments are dephased  randomly due to the white noise produced by the resistors. This can clearly be seen in Fig. \ref{fig:emi1} (a). Also, the normalized linewidth of $1.3\cdot10^{-4}$ is in very good agreement with the value calculated from $\Delta f_0$. Including back-action the line becomes sharper by about a factor of 2 for $M = 1$. By contrast, both for $M = 2$ and for $M = 10$ the linewidth is close to the case of zero back-action. 

%
%Figure Emi2 %%%%%%%%%%%%%%%%%%%%%%%%%%%%%%%%%%%%%%%%%%%%%%%%%%%%%%%%%%
\begin{figure}[tb]
\includegraphics[width=1\columnwidth,clip]{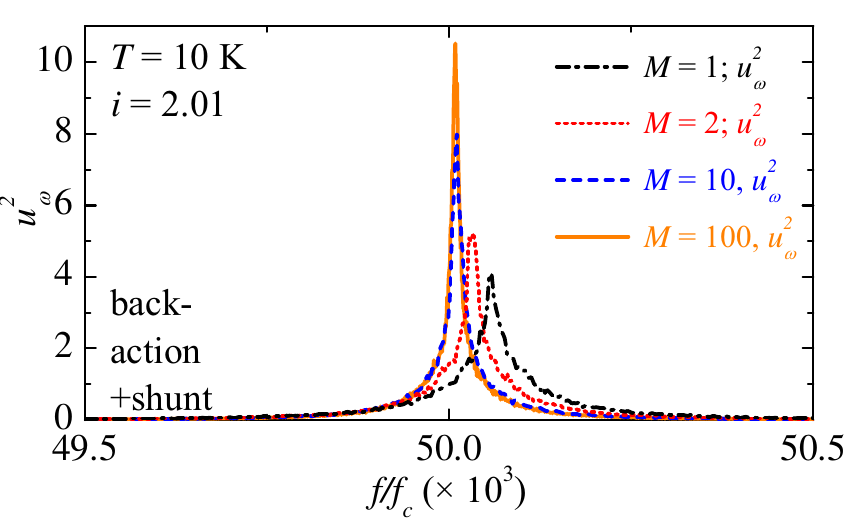}
\caption{(Color online) Fourier transforms (power) of the voltage $u$ across the stack vs. $f/f_c$ for $T_b$ = 10\,K,  $i = 2.01$ and $r_s/N = r_2$ for the thermally coupled circuit. 
}
\label{fig:emi2}
\end{figure}
%Figure Emi2 end %%%%%%%%%%%%%%%%%%%%%%%%%%%%%%%%%%%%%%%%%%%%%%%%%%%%%%
%
Fig. \ref{fig:emi2} shows corresponding data for the thermally coupled circuit in the presence of the shunt resistor. Note that in this plot the $u_\omega^2$ are \textit{not} multiplied with $M$. 
Obviously, $u_\omega^2$ increases with increasing $M$ indicating that phase-lock has occurred. In fact we have also checked this in a more traditional way by choosing different initial conditions for the Josephson phases of all segments; after some time these phases tended to approach the same value. The amplitudes $u_\omega^2$ increase (from 4 to 10.5) and $\Delta f$ decreases (from $3.7\cdot10^{-5}$ to $1.4\cdot10^{-5}$) roughly logarithmically with increasing $M$, i. e. a scaling $u_\omega^2 \propto M$ and $\Delta f\propto M$ is not observed. This indicates that the phase lock is not very strong at least for this bias point and for the large value of $\Gamma_0M/N$ used for the simulations.  
%
%Figure Emi3 %%%%%%%%%%%%%%%%%%%%%%%%%%%%%%%%%%%%%%%%%%%%%%%%%%%%%%%%%%
\begin{figure}[tb]
\includegraphics[width=1\columnwidth,clip]{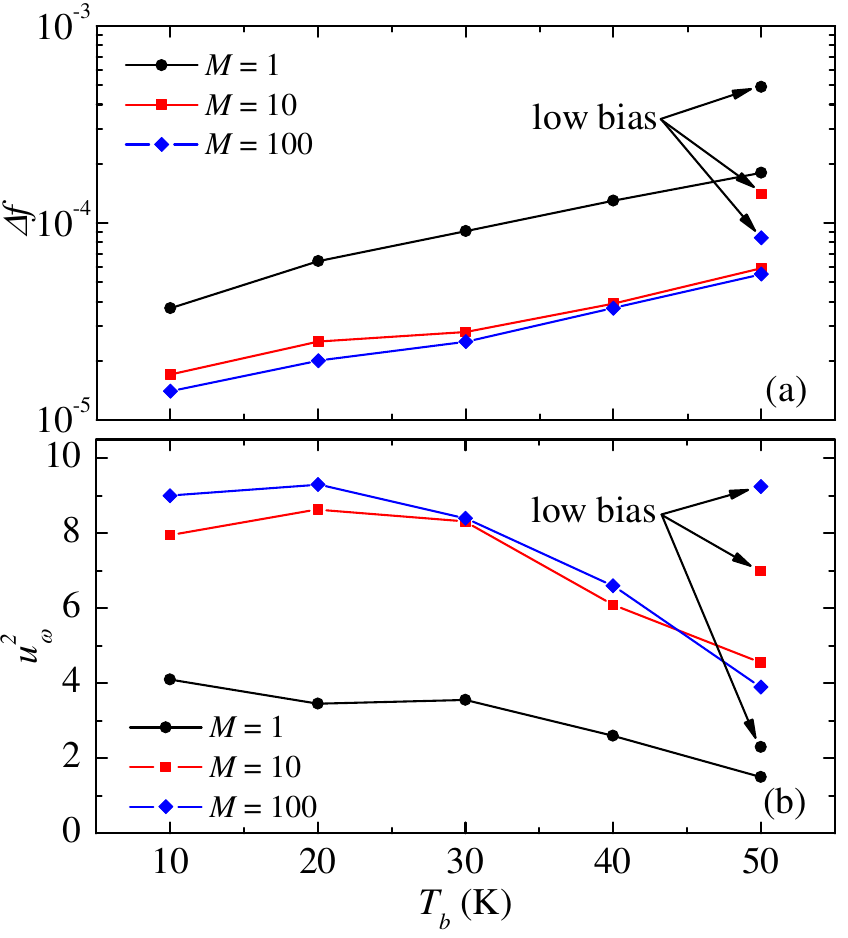}
\caption{(Color online) (a) Linewidth $\Delta f$ and (b) amplitude $u_\omega^2$ vs. bath temperature for the thermally coupled circuit with $r_s/N = r_2$ and the cases $M$ = 1, 10 and 100.  
}
\label{fig:emi3}
\end{figure}
%Figure Emi3 end %%%%%%%%%%%%%%%%%%%%%%%%%%%%%%%%%%%%%%%%%%%%%%%%%%%%%%
%
In Fig. \ref{fig:emi3}(a) we show $\Delta f$ vs. $T_b$ for the thermally coupled circuit including the shunt resistor. For the high bias data $\Delta f$ clearly increases with increasing $T_b$, i.e. the experimental observations are not reproduced at this level. Indeed we also performed similar calculations for other values of $v$ and obtained similar results. Further note the data points indicated by ``low bias'' in Fig. \ref{fig:emi3}. For all values of $M$ the linewidths taken at this bias are higher than the corresponding high-bias data points, although the differences get smaller with increasing $M$. Fig. \ref{fig:emi3}(b) shows the amplitude $u_\omega^2$ vs. $T_b$. For $T_b > 20$\,K  $u_\omega^2$ decreases with increasing bath temperature. For $M = 10$ and $M = 100$ the data points at 10\,K are somewhat lower than for 20\,K, indicating a shallow maximum near a bath temperature of 20\,K. Also this behavior is not in good agreement with measurements, where often the emission is maximum at intermediate temperatures between 30\,K and 40\,K, see e.g. Ref. \onlinecite{Wang10a}.

So far we have assumed that the parameters critical current and resistance are the same for all junctions. It has been emphasized however, that the finite slope of the edges of a BSCCO mesa leads to a gradient in these junction parameters \cite{Benseman11}. We account for this effect by introducing a linear increase of the junction critical currents via $i_{c,m} \propto (1+m\cdot a_{\rm{max}}/M)$ and a linear decrease of the resistances via $r_{k,m} \propto (1+m\cdot a_{\rm{max}}/M)^{-1}$. The parameter $a_{\rm{max}}$ controls the relative increase of the junction area between the bottom and the top of the stack and typically amounts to a few per cent in experiment. 
%
%Figure Emi4 %%%%%%%%%%%%%%%%%%%%%%%%%%%%%%%%%%%%%%%%%%%%%%%%%%%%%%%%%%
\begin{figure}[tb]
\includegraphics[width=1\columnwidth,clip]{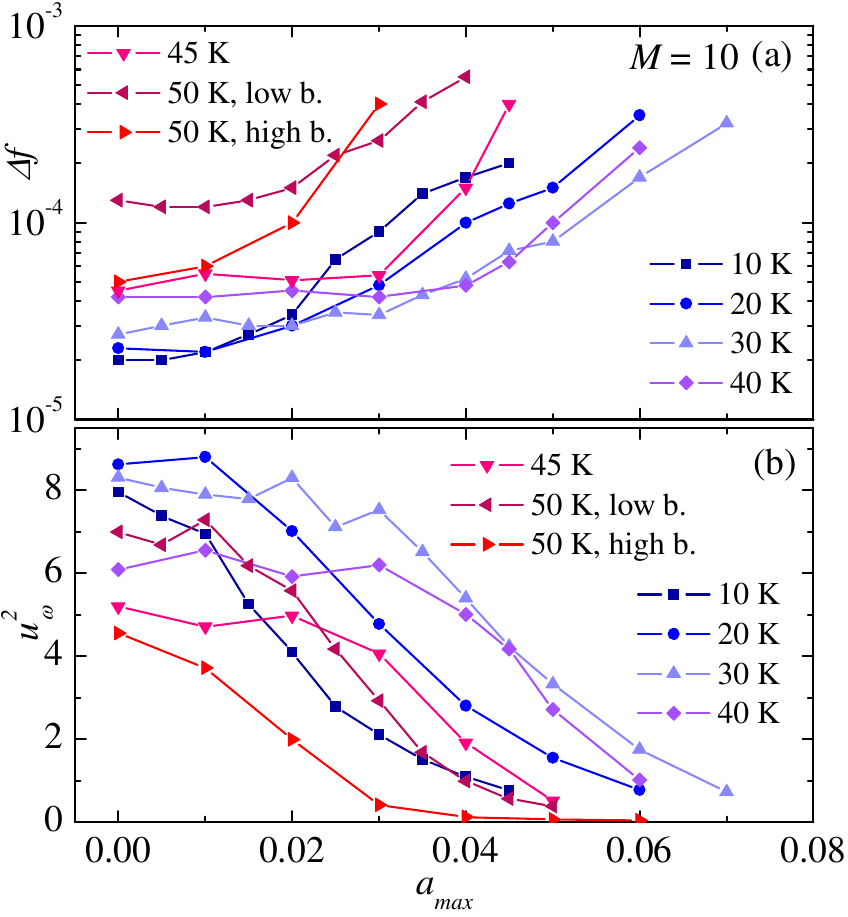}
\caption{(Color online) (a) Linewidth $\Delta f$ and (b) amplitude $u_\omega^2$ vs. relative maximum change in junction area $a_{\rm{max}}$ for various bath temperatures between 10 K and 50 K. $M = 10$.
}
\label{fig:emi4}
\end{figure}
%Figure Emi4 end %%%%%%%%%%%%%%%%%%%%%%%%%%%%%%%%%%%%%%%%%%%%%%%%%%%%%%
%
Fig. \ref{fig:emi4} shows $\Delta f$ and the amplitude $u_\omega^2$ vs. $a_{\rm{max}}$ for different bath temperatures. For a given $T_b$, $\Delta f$ increases and $u_\omega^2$ decreases with increasing $a_{\rm{max}}$. However, both the $\Delta f$ curves and the $u_\omega^2$ curves intersect for different values of $T_b$, showing that both $\Delta f$ vs. $T_b$ and $u_\omega^2$ vs. $T_b$ for fixed $a_{\rm{max}}$ can behave non-monotonously. In particular, for $a_{\rm{max}} > 0.02$ there are regimes where $\Delta f$ decreases with increasing bath temperature.  
%
%Figure Emi5 %%%%%%%%%%%%%%%%%%%%%%%%%%%%%%%%%%%%%%%%%%%%%%%%%%%%%%%%%%
\begin{figure}[tb]
\includegraphics[width=1\columnwidth,clip]{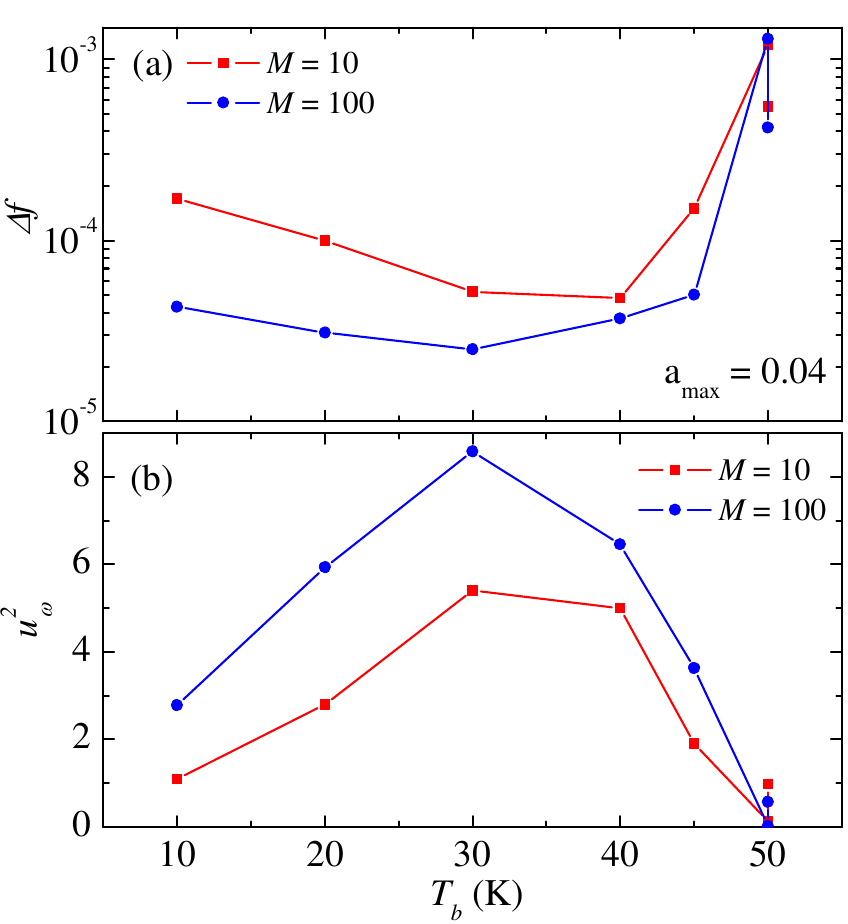}
\caption{(Color online) (a) Linewidth $\Delta f$ and (b) amplitude $u_\omega^2$ vs. $T_b$ for $M = 10$, $M = 100$ and $a_{\rm{max}} = 0.04$.
% (c) shows Fourier spectra $u_\omega^2$ vs. $f/f_c$ for various values of $T_b$. 
Terminating points at $T_b = 50$\,K in (a) and (b) are taken at low bias. 
}
\label{fig:emi5}
\end{figure}
%Figure Emi5 end %%%%%%%%%%%%%%%%%%%%%%%%%%%%%%%%%%%%%%%%%%%%%%%%%%%%%%
%

Fig. \ref{fig:emi5} shows this explicitly for $a_{\rm{max}} = 0.04$. $\Delta f$, cf. Fig. \ref{fig:emi5} (a), runs through a minimum, reached near $T_b = 40$\,K for $M = 10$ and near $T_b = 30$\,K for $M = 100$. Only here  $\Delta f$ is close to the value reached for $a_{\rm{max}} = 0$, c.f. Fig. \ref{fig:emi3}. The amplitude $u_\omega^2$ vs. $T_b$ runs through a pronounced maximum, similar as experimental data \cite{Wang10a}.
For  $a_{\rm{max}} = 0$ the decrease of $u_\omega^2$ at large values of  $T_b$ is essentially caused by the increase of thermal fluctuations. 
For $a_{\rm{max}} = 0.04$ this effect is present as well, leading to the decrease of $u_\omega^2$ at high temperatures. At low bath temperature the spread in junction parameters apparently affects $u_\omega^2$ strongly, causing the decrease of $u_\omega^2$ at low temperatures. 
We did not find a completely conclusive reason for this effect. However, it may have to do with an effective Stewart McCumber parameter
\begin{equation}
\label{Eq:bc_eff}
\beta_{c,\rm{eff}}=\frac{2\pi C I_c(T_b,i)R(T_b,i)^2}{\Phi_0},
\end{equation}
which governs the quality factor of the Josephson junctions at a given bias current $i$ and a given bath temperature $T_b$. This parameter should not be too low for good phase lock \cite{Hadley88}. At low values of $T_b$ the bias current is high ($i = 2.01$ at $T_b = 10$\,K) and both the ohmic resistance $v/i$ and the critical current $I_c(T_b,i)=I_{c1}(T_1)+I_{c1}(T_2)$ are low. At $T_b = 10$\,K and $i=2.01$ $\beta_{c,\rm{eff}}$ turns out to be about 0.06. For the bias points shown in Fig. \ref{fig:IVC} (b) $\beta_{c,\rm{eff}}$ increases monotonically with increasing $T_b$, reaching e.g. a value of 0.2 at $T_b$ = 50\,K and $i = 0.77$. 
%In any case the dependence of $\Delta f$ and also $u_\omega^2$ resembles the experimental data where, in addition to a decrease of  $\Delta f$ with increasing $T_b$, often a maximum in emission power is found at intermediate bath temperatures that are typically in the range 30-40\,K, see e.g. Ref. \onlinecite{Wang10a}.

For $f_{c0}$ = 7.5\,THz the normalized minimal linewidth of about $5\cdot10^{-5}$ ($2.5\cdot10^{-5}$), as calculated for $M = 10$ ($M = 100$), corresponds to a dimensioned linewidth of 370\,MHz (180\,MHz). Since we have taken a large value of $\Gamma_0 M/N$, not surprisingly this is larger than the smallest values of $\Delta f$ measured experimentally\cite{Li12}.
We thus finally also performed a simulation with $M=700$, using a more realistic value $\Gamma_0 = 10^{-5}$, and obtained a minimal linewidth of about 25\,MHz. This is in the range of the measured minimal linewidth.

We clearly emphasize  that we consider the model presented here just as a first step to describe the (Josephson) dynamics of stacked IJJs in the presence of heating. Nonetheless, the mechanism of phase synchronization via hot elements are likely to be present also in more sophisticated models. 
For example, coupled sine Gordon equations \cite{Sakai93,Kleiner94} could be combined with heat-diffusion equations in a simple manner as presented here. The implementation of such equations is  straightforward. However, calculating linewidths of radiation will be extremely time consuming, justifying the simplified approach taken in the present paper.  

\section{Conclusions}
In conclusion we have presented a simple model for intrinsic Josephson junctions stacks which are thermally coupled to a heat sink. The model incorporates two parallel arrays of Josephson junctions at temperatures $T_1$ and $T_2$ and an additional resistor at a temperature $T_2$ in parallel to the arrays.
The main motivation of our calculations was to provide a first step towards the description of terahertz dynamics of intrinsic junction stacks at high bias, where a hot spot coexists with a superconducting region. In experiment the emitted terahertz power is often found to be maximal at intermediate bath temperatures in the range 30--40\,K \cite{Wang10a}. Further, the linewidth of radiation decreases when $T_b$ is increased \cite{Li12}.  Both features are reproduced in our model, if a gradient in the junction parameters critical current and resistance is introduced. Such a gradient is likely to be present in experiment due to the sloped edges of IJJ stacks \cite{Benseman11}. It was found in our model that such a gradient leads to a larger degradation of phase locking properties at lower $T_b$ than at higher $T_b$. By contrast, thermal fluctuations, also degrading phase lock, increase with increasing $T_b$. These two effects counteract, leading to a maximum amplitude of the ac Josephson peak $u_\omega^2$  at an intermediate temperature near 40\,K and an increase of linewidth $\Delta f$ away from this maximum. In particular, the decrease of $\Delta f$ with increasing $T_b$ in the range 10\,K$ < T_b < $ 40\,K is reproduced qualitatively.
In spite of these encouraging results we strongly emphasize that we presented a zero order approach here. More sophisticated models like the 1D and 2D coupled sine-Gordon equations, with temperature dependent parameters, are clearly required to e.g. shine light on the interactions between the hotpot, cavity modes and linewidth of radiation. The present approach may show the way how to proceed in this direction.

%Acknowledgments
\acknowledgments
We gratefully acknowledge financial support by the JST/DFG strategic Japanese-German International Cooperative Program, the Grants-in-Aid for scientific research from JSPS, the National Natural Science Foundation of China (No.11234006), the Fundamental Research Funds for the Central Universities and Jiangsu Key Laboratory of Advanced Techniques for Manipulating Electromagnetic Waves, the RFBR and the Ministry of Education and Science of the Russian Federation.

%%%%end

%
%
%bib section starts here
\bibliography{hot-spots-waves_v2}
\end{document}